\documentclass[a4paper]{article}

\usepackage{geometry}
\usepackage{mathrsfs,amsthm,amsfonts}
\usepackage{mathtools}
\usepackage{framed}

\newtheorem{theorem}{\bf Theorem}[section]

\newtheorem{lemma}{\bf Lemma}[section]

\theoremstyle{definition}
\newtheorem{definition}{\bf Definition}[section]
\theoremstyle{remark}
\newtheorem{remark}{\bf Remark}[section]

\begin{document}

\title{Semi-martingale driven Variational Principles}

\author{
O. D. Street\thanks{Department of Mathematics, Imperial College, London, UK} \and D. Crisan$^{*}$}

\maketitle
\begin{abstract}
Spearheaded by the recent efforts to derive stochastic geophysical fluid dynamics models, e.g.,  \cite{cgh2017,ch2018, Holm2015,mt2016,m2014,rmc2015,rmc12017,rmc2017}, we present a generic framework for introducing stochasticity into variational principles through the concept of a semi-martingale driven variational principle and constraining the component variables to be compatible with the driving semi-martingale. Within this framework and the corresponding choice of constraints, the Euler-Poincare equation can be easily deduced. We show that their corresponding deterministic counterparts are particular cases of this class of stochastic variational principles.  Moreover, this is a natural framework that enables us to correctly characterize the pressure term in incompressible stochastic fluid models.  Other general constraints can also be incorporated as long as they are compatible with the driving semi-martingale.
\end{abstract}

\tableofcontents

\section{Introduction}

The characterisation of the dynamics of a physical system as a variational problem using the principle of stationary action is a classical problem that has received an enormous amount of attention. If we wish to explain a physical system by using a stochastic model, adding stochasticity at the level of the Lagrangian (the integrand of the action integral) has the benefit of preserving desirable qualities that the corresponding deterministic model may have, see \cite{Holm2015,Crisan2018,Crisan2019}. 
Guided by the tools of stochastic analysis, the purpose of this paper is to introduce a rigorous theoretical framework for adding stochasticity into a system through Hamilton's principle. This is achieved through a judicious choice of the class of models involved in Hamilton's principle.  Lagrange multipliers may be used to constrain a physical model to behave in a certain way as observed in the physical system of choice, e.g., incompressibility, the evolution of the advected quantities, restriction of motion to evolve on a given manifold, etc. These too can be chosen to be stochastic. Finally, the action function appearing in
Hamilton's principle can be defined as an integral with respect to a given measure which in the stochastic case can be chosen to be random. The model introduced below covers all these three options i.e., the classes of models, the Lagrange multipliers, as well as the integrator measure can all be chosen to be random as long as they remain compatible with an exogenously chosen semi-martingale. Subsequently, we will say that the corresponding variational principle is 
\emph{semi-martingale driven} (see definition \ref{def:SMDrivenVP}).

Stochastic modelling of fluid dynamics is rapidly growing in popularity due to the desire to model uncertainties in forecasts and improve data analysis methodologies. Fluid dynamics is a field of study in which variational principles hold particular interest. The application of variational principles to Eulerian fluid dynamics \cite{Seliger1968} was made possible by the consideration of Clebsch variables \cite{Clebsch1859}. This methodology provides a framework which can reveal profound information about the structure of a fluid equation. Indeed, V. Arnold's famous result about the geometric properties of the Euler equation \cite{Arnold1966} is an early example of precisely this. The development of these ideas have, in time, led to a unification of fluid systems using Euler-Poincar\'e theory \cite{HOLM19981}. This variational representation of fluid dynamics has presented an opportunity to consider probabilistic models which have some semblance to the structure of their deterministic cousins, and hence there is now an ever growing body of literature dedicated to exploiting this opportunity \cite{bismut1981,Lazaro-Cami2008,Lazaro-Cami2009,Kodama2014,Holm2018,kraus2019variational,Arnaudon2014,chen2015stochastic,Holm2015,Bou-Rabee2007,Bou-Rabee2009,Wang2007,Wang2009}. Whilst a stochastic action integral may be used to derive stochastic equations of motion, a similar approach may be used to derive deterministic equations which do not have a deterministic Hamiltonian structure. Namely, the Navier-Stokes equations have been derived via a stochastic version of the action corresponding to the Euler equations \cite{Inoue1979,Gomes2005,GregoryL.Eyink2010}. In the direction of deriving stochastic equations of motion, the idea of making use of such a structure when introducing randomness into a physical system first appeared in Hamiltonian mechanics \cite{bismut1981}. More recently, by building on the contributions of J.M. Bismut, stochastic Hamiltonian systems have been given a more rigorous treatment \cite{Lazaro-Cami2008,Lazaro-Cami2009}. Numerical integrators for systems of this type have also been considered \cite{Holm2018,kraus2019variational}

Whilst the Hamiltonian case has been considered carefully and in generality, the Lagrangian framework is less cohesively developed. Work in this direction has involved an Euler-Poincare\'e reduction featuring action integrals which take the form of an expected value of an integral \cite{Arnaudon2014}, as well as action integrals which are themselves stochastic integrals. This paper will be concerned with the latter. A number of papers have made use of action integrals of this form \cite{Bou-Rabee2007,Bou-Rabee2009,chen2015stochastic,Wang2007,Wang2009,Holm2015}, as well as considered their numerical integration \cite{Bou-Rabee2007,Bou-Rabee2009}. In particular, there are multiple ways of deriving stochastic partial differential equations for continuum dynamics with advected quantities \cite{chen2015stochastic,Holm2015}.
Although the framework we introduce encompasses a large class of possible stochastic variational principles, this work was inspired by and can be considered an extension of \cite{Holm2015}, where advected quantities are constrained to be advected according to a stochastic flow. As such, the examples we will use will be in the spirit of \cite{Holm2015}. We will place the equations derived in \cite{Holm2015} (and, by extension, those derived from similar variational principles) onto a more rigorous footing and properly formulate constraints imposed by Lagrange multipliers on equations of this type.  We detail the contribution of the paper in the following section.

\subsection{Contribution of the paper}

In this paper, we introduce a general framework for a large class of applications of Hamilton's principle to stochastic action integrals. We will discuss the necessary conditions which will ensure that our problem is properly formulated and makes sense from an analytical perspective. Stochastic models which result from such variational principles (in general) feature both deterministic and stochastic timestepping, i.e. the integral form of the equation of motion can feature integration with respect to both the Lebesgue measure `$dt$' as well as with respect to stochastic processes (for example Brownian motions). Through the introduction of a \emph{driving semi-martingale} we introduce a rigorous framework which encompasses all of the distinct types of integration occurring within the model. This helps us to avoid any potential complications which may occur when constraining the variational principle, since within this framework one need only ensure that each constraint is \emph{compatible} with the driving semi-martingale.

Significantly, in this work we will prove a stochastic version of the fundamental lemma of the calculus of variations (see Lemma \ref{lemma: StochasticFundamental}), which may be applied to any action integral which fits within our general framework. This ensures that the mechanics of how variations of stochastic integrals work is properly understood, and indeed it is a well defined mathematical process to undertake. From this result it is apparent that variations of stochastic integrals may be taken in an intuitive manner analogous to the deterministic picture, and this fact is now rigorously established.

Furthermore, we will consider the form of stochastic Lagrange multipliers. These will be formulated in such a way that constraints are compatible with the driving semi-martingale and thus constrain both the deterministic and stochastic parts of the dynamics. Pressure has the role of ensuring that the volume element remains constant and thus the flow is incompressible; in the case of stochastic fluid dynamics, the new framework highlights the need for the pressure to be thought of as a \emph{stochastic} Lagrange multiplier where time integration is considered with respect to each component of the driving semi-martingale. We will demonstrate this in the case of the Euler equations for incompressible fluids, and deduce an explicit stochastic differential equation for the pressure. At the level of the stochastic fluid equation, it is evident that the velocity formulation of the equation must feature both stochastic and deterministic pressure terms for the equation to make sense. However, this fact is not obvious from the variational principle unless one understands the proper form of the Lagrange multipliers which we formulate here. We believe that this is the first work where the stochastic equations of incompressible fluids, in velocity form, have been derived with a properly formulated pressure term from a stochastic action functional.

The introduction of stochasticity presented in this work generalizes the approach taken in  \cite{Holm2015}. In a forthcoming work, the driving semi-martingale is replaced by a rough path \cite{RoughPathsPreprint}, leading to the introduction of a new class of rough path driven variational principles.

\section{The General Case}

In this section, we will introduce a framework for the classical problem, through which the introduction of stochastic noise will be more easily understood. 

\subsection{The action function}

In classical mechanics, the \emph{action} of a physical system is given by the (deterministic) time integral of the corresponding Lagrangian, $L$. Its stationary points characterize the evolution 
of the physical system\footnote{We will, at times, focus our attention to the special case when we can consider this Lagrangian to be a spatial integral of some object, $\ell$ for example. Whilst $\ell$ can be thought of as a density for $L$, we will refer to both of these objects as Lagrangians throughout this paper.}.
If we denote the generalized coordinates of the system by $q(t)$, then the action $A(q)$ can be defined for any path $q(t)$
taken by the system as the time integral of the Lagrangian, $L$, along this path
\begin{equation}\label{TypicalAction}
    A(q)=\int_{t_0}^{t_1} L(t,q(t),\dot q(t))\,dt,
\end{equation}
where $\dot q(t)$ denotes a time derivative of $q$. This is the classical way of introducing the action function \cite{hand_finch_1998}. 
The above Lagrangian can be considered as a map $L: \mathbb{R}\times TQ\mapsto\mathbb{R}$, where $Q$ is the `configuration space' and $TQ$ denotes the corresponding tangent bundle, see \cite{HolmBook1,HolmBook2} for details. The framework we set up will accommodate time-dependent Lagrangians of this type, however we may at times consider Lagrangians which depend only on $TQ$. This framework is also applicable to reduced Lagrangians, which may be thought of as functionals on the Lie algebra corresponding to the configuration space, or even phase-space Lagrangians whose arguments are elements of the cotanget bundle of the configuration space. When setting up this framework we will therefore be purposely generic with regards to the space in which our Lagrangian is defined. However our examples will come from a particular case of Stochastic Advection by Lie Transport, which utilises reduced (Euler-Poincar\'e) Lagrangians. In section 3 we will thus transition towards Lagrangians of that type.

\emph{Hamilton's Principle} tells us that the physical path is distinct from the others in that it corresponds to a stationary value of the action function. Mathematically, Hamilton's Principle can be thought of as $\delta A =0$. Notice that this is equivalent to the value of the variational derivatives of $A$ being equal to zero with respect to each variable in the argument of $A$. In other words, Hamilton's Principle tells us that the variational derivatives, $\delta A(q)$ for example, of the action function corresponding to the physical path are null, see e.g. \cite{hand_finch_1998}. When modelling continuum media, the Lagrangian, $L$ takes the form of a spatial integral in a manner similar to that of field theory \cite{GOLEBIEWSKAHERRMANN1983300}. In this case, the action $A$ becomes an integral over both space and time. To be more precise, consider a space-time domain $\bar{\mathscr D}$ (with elements denoted by $\bar{x}$) and an action function given in the following compact form
\begin{equation}\label{GeneralActionIntegral}
	A(v):=\int_{\bar{\mathscr{D}}} \ell(\bar x, v(\bar x))\,d\bar\mu(\bar x), 
\end{equation}
where $\bar\mu(\bar x)$ is a measure on the given domain $\bar{\mathscr D}$ and $v$ encompasses all physical variables as well as the Lagrange multipliers used to constrain them. We can unravel \eqref{GeneralActionIntegral} in the following way: First, typically one separates $\bar{\mathscr{D}}$ into a time component and a space component and considers it to be a product space. Most commonly, this space can be chosen to be $[t_0,t_1]\times\mathscr{D}$ for $0\leq t_0 < t_2 \leq \infty$, where $\mathscr{D}\subset M$ is the spatial domain in which our dynamics occurs, and $M$ is a given manifold. Similarly, we ``unpack'' $\bar\mu$ in terms of the product between a measure $\mu$ on $\mathscr{D}$ and the Lebesgue measure for the time variable. The action integral can be written as
\begin{equation}\label{FamiliarContinuum}
	A(v):=\int_{t_0}^{t_1}\tilde\ell (t,v(t))\,dt, \ \ t_0,t_1\in [0,\infty),
\end{equation}
where
\begin{equation}\label{L}
	\tilde\ell (t,v(t)) = \int_\mathscr{D}\ell(t,x,v(t,x))\,\mu(dx).
\end{equation}
We recognize \eqref{FamiliarContinuum}+\eqref{L} as the typical action integral classically seen in continuum dynamics, (see, e.g. \cite{Arnold1966},\cite{HOLM19981},\cite{lahiri2005first}). In these equations, we recognise $\tilde\ell$ as the Lagrangian and $\ell$ as a Lagrangian density, see e.g. \cite{GOLEBIEWSKAHERRMANN1983300}. We will preserve this notation throughout the paper.

\begin{remark}
    Note that in \eqref{L}), we have used the notation $\mu(dx)$ to denote integration in the spatial variable $x\in \mathscr D$ with respect to the measure $\mu$. The alternative notation, $d\mu(x)$, where the `$d$' is positioned outside of the measure, will be reserved for the temporal integration. The reason for this will become clear in the following sections, where it is particularly necessary to distinguish between spatial and temporal integration.
\end{remark}

\subsection{Introducing stochasticity}\label{IntroductingStochasticity}

In the following,  we wish to  introduce stochasticity in the variational principle through the action function. Regardless of the source of stochasticity (through the choice of Lagrangian, the constraints and/or that of the dynamics), the action function will be assumed to be \emph{compatible} with a given semi-martingale $S$ (see Definition \ref{FunctionCompatibleDefinition} below for details). In other words, in the case of continuum dynamics, instead of satisfying \eqref{FamiliarContinuum}, we will consider $A(v)$ to be given by 
\begin{equation}\label{FamiliarContinuumS}
	A(v):=\int_{t_0}^{t_1}\tilde l (t,v(t))\circ d S_t,\quad t_0,t_1\in [0,\infty),
\end{equation}
where $\circ$ denotes the Stratonovitch integration (with respect to the semi-martingale $S$) and both $[t_0,t_1]\ni t \mapsto v$ and $[t_0,t_1]\ni t\mapsto \tilde l (t,v(t))$ will be assumed to be semi-martingales. This constraint will ensure that the integral in \eqref{FamiliarContinuumS} makes sense. 

The choice of the form \eqref{FamiliarContinuumS} for the stochastic version of the action function is justified by the following:

\begin{itemize}
    \item Formula  \eqref{FamiliarContinuumS} is a natural generalization of \eqref{FamiliarContinuum}. By choosing the semi-martingale to be given by $S_t\equiv t,\ \ t\in [t_0,t_1]$, the identity  \eqref{FamiliarContinuumS} reduces to  \eqref{FamiliarContinuum}.
    
    \item The stochastic calculus rules governing the Stratonovitch integral (as opposed to the It\^o integral) coincide with the classical (deterministic) calculus rules and therefore all the additional  technical details required by the introduction of stochasticity will be natural extensions of their deterministic counterparts.
    
    \item This generalization is a natural extension of the framework introduced in  \cite{Holm2015}, where the driving semi-martingale is given by $S_t=(t, W^1_t,...,W^n_t,...)$. In other words, in \cite{Holm2015}, $S_t$ is an infinite dimensional stochastic process with the first component identically equal with the time variable and the rest of the components being given by independent Brownian motions. In particular it covers models where the advected quantities are constrained to follow stochastically perturbed trajectories. See section $\ref{SALTSection}$ below for details.
    
    \item By a judicious choice of the driving semi-martingale one can introduce non-independent noise increments (for example through an Ornstein-Uhlenbeck process) in order to incorporate memory into the fluid dynamic model \cite{Holm_2018}.
    
    \item The new framework lends itself easily to extensions to  non-flat spaces (manifolds), where the It\^o based stochastic calculus does not have an intrinsic development, see \cite{Hsu_book}.
    
    \item It is a natural precursor of a new class of rough path driven variational principles, see \cite{RoughPathsPreprint}.
    
\end{itemize}

In the following we will formalize a stochastic generalization the action integral of the type \eqref{TypicalAction}, with some Lagrangian, $L(t,x,v(t,x))$. In order to do so, we must define a number of mathematical objects.

    Let $(\Omega,\mathcal{F},P)$ be a probability space endowed with a 
    filtration  $\{\mathcal{F}_t\}_{t\ge 0}$ that satisfies the usual 
    conditions and $B$ be an arbitrary Banach space. A $B$-valued $\{\mathcal{F}_t\}_{t\ge 0}$-adapted stochastic process $X$ is a family of random variables $(X_t)_{t\ge0}$, where $X_t:\Omega\mapsto B$, parameterized by $t\ge 0$ such that for all $t \ge 0,\,X_t$ is $\mathcal{F}_t$ measurable.
All standard notions of basic probability theory (integration, conditional expectation, etc) are easily extended to $B$-valued random variables and $B$-valued random processes. Similarly, all standard notions of stochastic calculus are easily extended from (finite dimensional) Euclidean spaces to $B$-valued  stochastic processes. In order to generalize our action integral we will consider function valued stochastic processes, which can be thought of as a specific class of $B$-valued stochastic processes.

Recall that $\mathscr{D}$ denotes the spatial domain in which our dynamics is occurring. We denote by $\cal F(\mathscr{D})$ a suitably chosen space of functions over $\mathscr{D}$ equipped with a norm $\|\cdot \|_{\cal F(\mathscr{D})}$ so that the normed space $\cal F(\mathscr{D})$ is indeed a Banach space, and ${\cal F(\mathscr{D})}^n$ denotes a similar space of $n$-dimensional versions of these functions.\footnote{By ${\cal F(\mathscr{D})}^n$ we mean that the functions themselves have $n$ components, not that the function is defined on an $n$ dimensional domain.} We will define what it means for a semi-martingale to be "compatible" with respect to a classical $\mathbb{R}^\mathbb{N}$-valued semi-martingale.

\begin{remark}
    We will use the notation `$\circ$' to denote Fisk-Stratonovich integration with respect to a semi-martingale,\footnote{See e.g. \cite{Karatzas1998} for further details regarding the Fisk-Stratonovich integration.} which will frequently appear when we integrate in the time variable.\footnote{The notation `$\circ$' may also be used to denote the composition of functions but it will be clear from the context when this is the case.}
\end{remark}

\begin{definition}\label{FunctionCompatibleDefinition}
An ${\cal F(\mathscr{D})}^n$-valued semi-martingale, $g_t$, is called  compatible with respect to a continuous $\mathbb{R}^\mathbb{N}$-valued semi-martingale, $S_t=\{S_t^j,\ j\geq 1\}$ if there exists a set of ${\cal F(\mathscr{D})}$-valued continuous semi-martingales $G_t = \{G_t^{i,j}: i=1,\dots, n,\ j=1,2,\dots \}$ such that
\begin{equation}\label{FunctionCompatibleEquation}
    g_t^i = g_0^i + \sum_j\int_0^t G_s^{i,j} \circ dS_s^j,
    \ \ \ i=1,\dots,n,
\end{equation}
and the semi-martingales $G^{i,j}$ are chosen so that the infinite sums in \eqref{FunctionCompatibleEquation} make sense.
The system of identities \eqref{FunctionCompatibleEquation} is written in integral form componentwise, and can be compactly re-written in differential form as 
\begin{equation}\label{FunctionCompatibleEquationReduced}
    \text{d}g_t = G_t\circ dS_t,
\end{equation}
where the above equation encompasses all the relevant summation and all individual components. Note that the continuous semi-martingales $G_t^{i,j}$  do not necessarily need to be compatible with respect to $S_t$. We say that $G_t$ represents the \emph{stochastic derivative} of $g$ with respect to the driving semi-martingale $S_t$.
\end{definition}

\begin{remark}
    We will consistently use the above lower and upper case notation as in \eqref{FunctionCompatibleEquation} and \eqref{FunctionCompatibleEquationReduced} for objects which are compatible with $S_t$.
\end{remark}

\begin{remark}\label{IntegrabilityRemark}
   In order to make sense of the infinite sums in \eqref{FunctionCompatibleEquationReduced}, we will need to impose constraints on the choice of the semi-martingales $G^{i,j}$, $j=1,2,\dots$ $i=1,\dots, n$,. Let us identify the finite variation parts and the martingale parts of $S^{j}$ $j=1,2,\dots$ as
   \begin{equation}\label{S_DoobMeyer}
       S^j = B^j + M^j,
   \end{equation}
   where $B^j$ and $M^j$ are the finite variation and martingale parts of $S^j$, respectively\footnote{\eqref{S_DoobMeyer} is the Doob-Meyer decomposition of $S^j$.}. We will assume that $G^{i,j}$ will be integrable with respect to $B^j$ for all $j\geq 1$ and that
   \begin{equation}\label{FiniteVariationIntegralCondition}
       \mathbb{E}\left[\left( \sum_{i=1}^\infty \int_0^t \|G^{i,j}\|_{\cal F(\mathscr{D})}\, dV_{B^j} \right)^2\right] < \infty,
   \end{equation}
   where $V_{B^j}$ is the variation process corresponding to $B^j$. Separately we will assume that
   \begin{equation}\label{MartingaleIntegralCondition}
       \mathbb{E}\left[\sum_{i=1}^\infty \int_0^t \|G^{i,j}_s \|^2_{\cal F(\mathscr{D})}\, d[M^j]_s \right] < \infty,
   \end{equation}
   where $[M^j]$ is the quadratic variation of the martingale $M^j$. We have that \eqref{FiniteVariationIntegralCondition} and \eqref{MartingaleIntegralCondition} together with the assumption that $\|g_0^i\|_{\cal F(\mathscr{D})}<\infty$ implies that the semi-martingales $g_s^i$ are well defined and are square integrable. More precisely
   \begin{equation*}
       \mathbb{E}\left[ \sup_{s\in[0,t]}\|g^i\|_{\cal F(\mathscr{D})} \right] < \infty.
   \end{equation*}
   We will henceforth refer to conditions such as these as `integrability constraints' on $G^{i,j}$, since we will need similar conditions on other objects.
\end{remark}

\begin{remark}
    In the following we can assume, without loss of generality, that all the finite variation terms of the semi-martingale $S_t$ are collected into one of its components, with all other components consisting of (continuous) martingales. 
\end{remark}

\begin{definition}
    The continuous process $S_t$ in Definition \ref{FunctionCompatibleDefinition} will be called the \emph{driving semi-martingale} of the system.
\end{definition}

Suppose $g_t$ and $S_t$ are as in Definition \ref{FunctionCompatibleDefinition}, then for each $i$ and $j$,
\[
t\rightarrow \int_0^t g_t^j \circ dS_t^i,
\]
is a well defined square-integrable one-dimensional semi-martingale provided $g_t^j$ satisfies similar integrability conditions to those presented in Remark \ref{IntegrabilityRemark}.

We are now in a position to define the stochastic generalization of the action \eqref{TypicalAction}. Suppose we have a driving semi-martingale, $S_t=\{S_t^i,i\ge 1 \}$ and assume that the process $v(t)$ that models all physical variables and all the Lagrange multipliers is a continuous semi-martingale which is compatible with $S_t$ (see Remark \ref{MultiplierCompatibilityRemark}). Let the Lagrangian $\tilde\ell$ be a $\mathbb{R}^\mathbb{N}$-valued semi-martingale such that the process $t\mapsto \tilde\ell(t,v(t)), \ \ t\in [t_0,t_1]$ is compatible with $S_t$, then
\begin{equation}\label{StochasticTypicalAction}
    \int_{t_0}^{t_1}\tilde\ell(t,v(t))\circ dS_t,
\end{equation}
is a well defined square integrable one-dimensional semi-martingale.\footnote{Note that we intrinsically assume here that $\tilde\ell$ satisfies similar integrability conditions as specified for $G^{i,j}$.}
\begin{definition}\label{def:SMDrivenVP}
    By a \emph{semi-martingale driven variational principle}, we mean the application of the principle of stationary action to a well defined stochastic action integral of the form \eqref{StochasticTypicalAction}
    in order to derive the corresponding stochastic governing equation for the chosen physical system.
\end{definition}

As we will see in section \ref{StochasticVariationalDerivativesSection}, the requirement that the components of $S_t$ are orthogonal will ensure that we may apply Hamilton's Principle and the variational derivatives operate in a way which is analogous to the deterministic picture. 

\begin{remark}
    If the driving semi-martingale, $S_t$, is identically equal to the time variable $S_t=t, t\in [t_0,t_1]$, then our system reverts back to the deterministic picture described in the previous section.
\end{remark}

\subsection{Stochastic continuum dynamics}\label{StochasticContinuumDynamicsSection}

In the special case of continuum dynamics, we wish to define a stochastic generalization of the action integrals \eqref{GeneralActionIntegral} and \eqref{FamiliarContinuum}. In the case where our spatial integral is a standard deterministic integral (i.e. the measure $\mu$ is the Lebesgue measure on $\mathscr D$ , we do not need any extra equipment other than that presented in Section \ref{IntroductingStochasticity} and we may hide the deterministic spatial integration inside the Lagrangian $\ell$. However, to prepare the basis for a more general framework, we will make use of an additional class of measure valued stochastic processes (these will be used to replace the measure $\bar\mu$ in \eqref{GeneralActionIntegral}) as well as the function valued stochastic processes described in section \ref{IntroductingStochasticity}

Let $\cal M(\mathscr{D})$ be the space of finite measures\footnote{The methodology presented here can be easily extended to spaces of non-finite measures (e.g. $\sigma$-finite measures such as the Lebesgue measure on the $d$-dimensional Euclidean space.}  over $\mathscr{D}$ endowed with the total variation norm and ${\cal{M}(\mathscr{D})}^n$ denote the space of $n$-dimensional versions of these. In the following we will work with ${\cal{M}(\mathscr{D})}^n$-valued semi-martingales, see, e.g. \cite{Kal2017} for a characterization of this class of stochastic processes.  

\begin{definition}\label{MeasureCompatibleDefinition}
    An ${{\cal M}(\mathscr{D})}^n$-valued semi-martingale $\nu_t$, is called compatible with respect to a given $\mathbb{R}^\mathbb{N}$-valued semi-martingale, $S_t$, if $\nu_t=(\nu_t^i)_{i=1}^n$ has a representation of the form
    \begin{equation}\label{MeasureCompatibleEquation}
        \nu_t^i=\nu_0^i+\sum_j\int_0^t \mu_s^{i,j}\circ dS_s^j,
        \ \ \ i=1,\dots,n,
    \end{equation}
where $\mu_t^{i,j}$ are continuous  ${{\cal M}(\mathscr{D})}$-valued semi-martingales for every $i=1,\dots,n, j=1,2,\dots$. Similarly to in definition \ref{FunctionCompatibleDefinition}, the system of equations \eqref{MeasureCompatibleEquation} can be re-written in the following compact form:
\begin{equation}
    \text{d}\nu_t = \mu_s\circ dS_s.
\end{equation}
\end{definition}

\begin{remark}
   Similar integrability constraints to those imposed on $G^{i,j}$ in Remark \ref{IntegrabilityRemark} are needed here to ensure that the $\nu_i$ are well defined.
\end{remark}

 Now suppose that $g,G,\mu,\text{and }\nu$ are as defined in Definitions \ref{FunctionCompatibleDefinition} and \ref{MeasureCompatibleDefinition}, then we have
\[
\int_{\mathscr D} g_t^i (x)\nu_t^i(dx)=\int_{\mathscr D} g_0 ^i(x)\nu_0^i(dx) + \sum_j\int_0^t \int_{\mathscr D} G_t^{i,j} \nu_t^i(dx) \circ dS_t^j+\sum_j\int_0^t \int_{\mathscr D} g_s^i\mu_s^{i,j}(dx)\circ dS_t^j,
\]
where we have assumed that the relevant integrability conditions are satisfied for this to be well defined. We can think of this as a stochastic version of the chain rule
\[
d (g_t^i \nu_t^i)= (dg_t^i) \nu_t^i+ g_t^i d(\nu_t^i).
\]

If Definitions \ref{FunctionCompatibleDefinition} and \ref{MeasureCompatibleDefinition} hold, then it follows that the processes 
\begin{equation}\label{ContinuumActionProcess}
    t\rightarrow \int_0^t\int_{\cal D} g_t(x)^i\circ d\nu_t^i(dx),
\end{equation}
and
\begin{equation}
    t\rightarrow \int_0^t\int_{\cal D} g_t(x)^i\mu_t^{i,j}(dx)\circ dS_t^j,
\end{equation}
are well defined one-dimensional semi-martingales. Notice that equivalence of the above processes is a direct consequence of Definition \ref{MeasureCompatibleDefinition}, which governs the way in which we define our time integration as being with respect to $S_t$. Again we assume here that the integrability constraints are satisfied in order for these integrals to make sense.
\begin{remark}
    In equation \eqref{ContinuumActionProcess}, the spatial integration is defined by integrating with respect to $\nu$ as a measure. Recall that we denote spatial integration by writing `$dx$' as the argument of the measure $\nu$. The temporal integration is achieved by integrating with respect to $\nu_t$ as a stochastic process, which is well defined because of its compatibility with respect to the driving semi-martingale.
\end{remark}

As we have mentioned before, by $v$ we mean the collection of all physical variables and Lagrange multipliers. In order to generalize our action for continuum dynamics, we now clarify the spaces in which these objects live. Let $\mathfrak{X}=\mathfrak{X}(M)$ denote the space of smooth vector fields on our manifold $M$ with appropriate boundary conditions on the boundary $\partial\mathscr{D}$ of our spatial domain. We define the non-degenerate $L^2$ pairing between $\mathfrak{X}$ and its dual space $\mathfrak{X}^*$ by \[\langle\cdot,\cdot\rangle_\mathfrak{X}:\mathfrak{X}^*\times\mathfrak{X}\to\mathbb{R}.\] Furthermore, let $V=V(M)$ be a vector space which contains the geometric quantities that typically occur in ideal continuum dynamics \cite{Holm2015}. This includes scalar functions and $k$-forms in all dimensions, all of which we assume to be sufficiently smooth. We again define a non-degenerate $L^2$ pairing between this and its dual by \[\langle\cdot,\cdot\rangle_V:V^*\times V\to\mathbb{R}.\]
We will choose the components of $v$ to lie within $\mathfrak{X}$, $V$, $\mathfrak{X}^*$, or $V^*$ depending on what they are. In other words, $v\in \prod \mathfrak{B}^i = \mathfrak{B}$, where $\mathfrak{B}^i = \mathfrak{X}$, $V$, $\mathfrak{X}^*$, or $V^*$. For an example of these spaces, see the remark following Theorem \ref{EulerPoincareTheorem}.

Suppose we have a $\mathbb{R}^\mathbb{N}$-valued driving semi-martingale $S_t$, a ${{\cal M}(\mathscr D)}^\mathbb{N}$-valued semi-martingale $\nu_t$ which is compatible with $S_t$. Let the components of $v$ be $\mathfrak{B}$-valued semi-martingales which are compatible with $S_t$. Suppose also that we have a Lagrangian $\ell$ which is such that $t\mapsto\ell(t,x,v(t,x))$ is a $\mathbb{R}^\mathbb{N}$-valued semi-martingale compatible with $S_t$. We then have a well defined stochastic action integral given by
    \begin{equation}\label{GeneralActionIntegralStochastic}
        \int_{t_0}^{t_1}\int_\mathscr{D} \ell(t,x,v(t,x))\circ d\nu_t(dx),
    \end{equation}
and we can look to derive the corresponding stochastic governing equation via the semi-martingale driven variational principle.

Note that \eqref{GeneralActionIntegralStochastic} can be thought of as a stochastic generalization of the deterministic general action integral \eqref{GeneralActionIntegral} 
and the deterministic familiar action integral \eqref{FamiliarContinuum} becomes
\begin{equation}\label{FamiliarContinuumStochastic}
	\int_{t_0}^{t_1}\int_{\mathscr{D}} \ell(t,x,v(t,x))\,\mu_t(dx)\circ dS_t,
\end{equation}
in the stochastic framework. The crucial difference between the stochastic and deterministic action integrals is that all time integration is now stochastic rather than integration with respect to the Lebesgue measure, `$dt$'.

\subsection{Comparison between the stochastic and deterministic frameworks}\label{StochasticDeterministicComparison}

Introducing stochasticity into a mathematical model generates complexities into components of the model which do not occur in the deterministic case. For a differentiable path $t\mapsto v(t)$, the object $\text{d}v$ can be thought of as the stochastic version of the object `$\dot v\,dt$', where $\dot v$ is the time derivative of $v$, which appears in the deterministic framework. This can loosely be thought of as a variable together with the measure with respect to which we are integrating. In the deterministic case, this is an entirely trivial object since the `$dt$' integration is standard (Lebesgue integration). Furthermore, deterministic equations tend to be written in differential form rather than in an integral form. In the stochastic case, the variables are represented as paths $[t_0,t_1]\ni t\mapsto v(t,\cdot)\in\mathfrak{B}$ which are not classically differentiable in the time component, thus it is a necessity to write our equations in integral form. It is essential that we take care with how we define the integrals and what object we integrate with respect to. The framework we present here provides a clear methodology to easily understand how integration must be defined.

As a sanity check to ensure that we are creating a framework compatible to the classical deterministic one, consider the special case where the driving semi-martingale is $ S_t=t$. Notice that in this case we recover exactly the deterministic picture. In particular, the action integral \eqref{FamiliarContinuumStochastic} reduces to \eqref{FamiliarContinuum}. Similarly, the compatibility condition reduces to the assumption that the process $v$ can be written as
\[
    v_t = v_0 + \int_0^t V_s\,ds 
\]
where $V$ is continuous, which is equivalent to requiring that  $\frac{\partial v}{\partial t} = V$, in other words that our variables are differentiable in time and have continuous derivatives. 

\subsection{Variational calculus and stochastic integrals}\label{StochasticVariationalDerivativesSection}

For a given action, $A$, Hamilton's Principle can be mathematically considered to be the statement that the first variation of $A$ is zero: \[\delta A =0. \] We want to verify that if our action is defined to be a stochastic integral, then we may take variations similarly to the deterministic case. Note that the first variation is precisely defined via the G\^ateaux variation.

Suppose that $A$ has the form
\[
A(v) = \int_{t_0}^{t_1}\tilde\ell (t,v(t))\circ dS_t,\quad t_0,t_1\in [0,\infty),
\]
where $v\in\mathfrak{B}$. Then $\delta v \in \mathfrak{B}$, $\delta \tilde \ell / \delta v \in \mathfrak{B}^*$, and  by definition we have
\[
0 = \delta A(v) = \int_{t_0}^{t_1} \delta \tilde \ell (t,v(t)) \circ dS_t = \int_{t_0}^{t_1} \left\langle \frac{\delta\tilde\ell}{\delta v},\delta v \right\rangle_{\mathfrak{B}}\circ dS_t = \int_{t_0}^{t_1}\int_\mathscr{D} \frac{\delta\tilde\ell}{\delta v}\delta v \,\mu(dx)\circ dS_t.
\]
Recall that we say $\tilde\ell$ is $\mathbb{N}$ dimensional, meaning that its dimension is equivalent to that of the driving semi-martingale $S_t$. In other words, it has precisely the required amount of components such that the action integral could be re-written as
\[
A(v) = \int_{t_0}^{t_1}\sum_i\tilde\ell^i(t,v(t))\circ dS_t^i.
\]
\begin{lemma}[Stochastic fundamental lemma of the calculus of variations]\label{lemma: StochasticFundamental}
	Suppose $f(t,x)$ is a ${\cal F(\mathscr{D})^\mathbb{N}}$-valued semi-martingale. If for any ${\cal F(\mathscr{D})^\mathbb{N}}$-valued semi-martingale, $\psi(t,x)$, we have 
	\begin{equation}\label{LemmaAssumption}
	\int_{t_0}^{t_1}\int_\mathscr{D}f(t,x)\psi(t,x) \,\mu(dx)\circ dS_t = 0,
	\end{equation}
	then for any $\alpha,\ \beta$ such that $t_0\leq \alpha < \beta \leq t_1$ we have
	\begin{equation}\label{LemmaConclusion}
	\int_\alpha^\beta f(t,x)\circ dS_t = 0.
	\end{equation}
\end{lemma}

\begin{remark}\label{remark:VariationalDerivatives}
In the context of our action integral, equation \eqref{LemmaConclusion} tells us that
\begin{equation}\label{VariationalDerivativesSPDE}
    \int_\alpha^\beta \frac{\delta\tilde\ell}{\delta v} \circ dS_t=\int_\alpha^\beta\sum_i \frac{\delta\tilde\ell^i}{\delta v} \circ dS_t^i = 0,
\end{equation}
which in practice can often be written as a SPDE, as we will see in the proof of Theorem \ref{EulerPoincareTheorem}.
\end{remark}

\begin{proof}
    Since \eqref{LemmaAssumption} holds for any such $\psi$, it holds for semi-martingales of the form
    \begin{equation*}
        \psi(t,x) = (\psi^0(t,x),\psi^1(t,x),\dots)= (\phi(x)\varphi^0(t),\phi(x)\varphi^1(t),\dots) = \phi(x)\varphi(t),
    \end{equation*}
    where $\varphi:[t_0,t_1]\to \mathbb{R}^\mathbb{N}$ is a smooth function such that $\varphi(t_0)=\varphi(t_1)=0$ and $\phi:\mathscr{D}\to\mathbb{R}$ belongs to a class of functions $\mathscr{S}$ which represents a total set. By this we mean that the class of functions $\mathscr{S}$ is such that if, for some function $g:\mathscr{D}\to\mathbb{R}$, we have
    \begin{equation*}
        \int_\mathscr{D} g(x)\phi(x)\,\mu(dx),\quad\forall\,\phi\in\mathscr{S}
    \end{equation*}
    then $\mu$ almost surely we have that $g=0$.
    
    We now define $F_\phi:[t_0,t_1]\to\mathbb{R}^\mathbb{N}$ by
    \begin{equation*}
        F_\phi(t) = \int_\mathscr{D} f(t,x)\phi(x)\,\mu(dx).
    \end{equation*}
    Then $F_\phi$ is a semi-martingale and moreover the covariation process $[F_\phi,S]_t$ is well defined. In particular,
    \begin{equation*}
        [F_\phi,S]_t = \int_\mathscr{D} [f,S]_t(x)\phi(x)\,\mu(dx).
    \end{equation*}
    Since $\varphi$ is smooth, $F_\phi\varphi$ is also a semi-martingale and $[F_\phi\varphi , S]_t$ is well defined with
    \begin{equation*}
        [F_\phi\varphi,S]_t = \int_{t_0}^{t_1} \varphi(t)\,d[F_\phi,S]_t.
    \end{equation*}
    It follows that
    \begin{align*}
        0 &= \int_{t_0}^{t_1}\int_\mathscr{D}f(t,x)\psi(t,x)\,\mu(dx)\circ dS_t = \int_{t_0}^{t_1}F_\phi(t)\varphi(t)\circ dS_t \\
        &= \int_{t_0}^{t_1}F_\phi(t)\varphi(t)\,dS_t + \frac{1}{2}\int_{t_0}^{t_1}F_\phi(t)\varphi(t)\,d[F_\phi,S]_t.
    \end{align*}
    For arbitrary $\alpha$ and $\beta$ such that $t_0\leq \alpha < \beta \leq t_1$, choose $\varphi = \mathbb{1}_{[\alpha,\beta]}$ where $\mathbb{1}$ is the indicator function. Let $(\varphi_n)_{n=1}^\infty$ be a uniformly bounded sequence of smooth functions such that
    \begin{equation*}
        \varphi_n \to \mathbb{1}_{[\alpha,\beta]},
    \end{equation*}
    where this convergence is pointwise. Then, by the It\^o isometry and the bounded convergence theorem, we have
    \begin{equation*}
        \mathbb{E}\left[\left(\int_{t_0}^{t_1}F_\phi(t)(\varphi_n(t)-\mathbb{1}_{[\alpha,\beta]}(t))dS_t\right)^2\right] = \mathbb{E}\left[\int_{t_0}^{t_1}\big(F_\phi(t)(\varphi_n(t)-\mathbb{1}_{[\alpha,\beta]}(t))\big)^2\,d[S]_t\right] \to 0,
    \end{equation*}
    and
    \begin{equation*}
        \int_{t_0}^{t_1}|F_\phi(t)||\varphi_n(t)-\mathbb{1}_{[\alpha,\beta]}(t)|\,d[F_\phi,S]_t \to 0.
    \end{equation*}
    Therefore, we have
    \begin{align*}
        0 &= \int_{t_0}^{t_1}F_\phi(t)\varphi_n(t)\,dS_t + \frac{1}{2}\int_{t_0}^{t_1}F_\phi(t)\varphi_n(t)\,d[F_\phi,S]_t \\
        &\to \int_{t_0}^{t_1}F_\phi(t)\mathbb{1}_{[\alpha,\beta]}(t)\,dS_t + \frac{1}{2}\int_{t_0}^{t_1}F_\phi(t)\mathbb{1}_{[\alpha,\beta]}(t)\,d[F_\phi,S]_t = \int_\alpha^\beta F_\phi(t)\circ dS_t.
    \end{align*}
   The stochastic Fubini theorem then gives
   \begin{align*}
       0 &= \int_\alpha^\beta F_\phi(t)\circ dS_t = \int_\alpha^\beta \int_\mathscr{D}f(t,x)\phi(x)\,\mu(dx)\circ dS_t \\
       &= \int_\mathscr{D}\int_\alpha^\beta f(t,x)\phi(x)\circ dS_t\,\mu(dx) = \int_\mathscr{D}\left(\int_\alpha^\beta f(t,x)\circ dS_t\right)\phi(x)\,\mu(dx),
   \end{align*}
   and the total set property then gives our result.
    
\end{proof}

We thus notice that taking variations with respect to the stochastic action integral is analogous to the deterministic case, and operationally can be performed in a similar manner by the observation made in Remark \ref{remark:VariationalDerivatives}.

\begin{remark}
   Should we wish to consider \eqref{VariationalDerivativesSPDE} componentwise, then we will need to place additional assumptions on the driving semi-martingale $S_t$. Once we do this, we can obtain relations $\delta\tilde l^i / \delta v = 0$, $\forall i$. Whilst this may feel more analogous to the deterministic case, in our examples we will see that, in practice, the components of $\tilde \ell$ may have little to no physical meaning. 
\end{remark}

\subsection{Stochastic Lagrange multipliers}

Given some Lagrangian, we want to to make rigorous sense of any constraints in the form of Lagrange multipliers which this may contain. Recall that $v$ denotes all physical variables as well as any Lagrange multipliers. We will now decompose $v$ into $\tilde v$ and $\Lambda$, where $\tilde v$ are the physical variables and $\Lambda$ are the Lagrange multipliers. Note that this identification is sometimes artificial; for example, when a Lagrange multiplier enforces a dynamic constraint (in the deterministic case, this corresponds to a constraint featuring time derivatives), the Lagrange multiplier itself can have an evolution equation and can have an interpretation as a physical variable. Let us now formulate explicitly how to use Lagrange multipliers to constrain our system. We will denote by $\tilde\ell(t,x,\tilde v(t,x))$ the unconstrained Lagrangian.

In our action integral, we can enforce a constraint on our variables by defining some function $C = C(t,x,\tilde v(t,x))$ in $\mathfrak{B}$, and incorporate this into the action integral together with a Lagrange multiplier. The action \eqref{StochasticTypicalAction} with Lagrange multipliers takes the form
\begin{equation*}
	\int_{t_0}^{t_1} \tilde\ell(t,x,\tilde v(t,x)) - \langle \Lambda, C(t,x,\tilde v(t,x)) \rangle_\mathfrak{B} \circ dS_t,
\end{equation*}
where $\Lambda \in \mathfrak{B}^*$ is the Lagrange multiplier. We can think of the time integration as being attached to either the constraint or the Lagrange multiplier, i.e. the inner product can be written as $\langle \Lambda, C(t,x,\tilde v(t,x))\circ dS_t\rangle$ or $\langle \Lambda\circ dS_t, C(t,x,\tilde v(t,x))\rangle$ and when incorporating either of these into the above action integral, we define the resulting integral to be the same as the above. In the former case, the notation can be condensed by defining a function-valued semi-martingale, $\lambda$, which is compatible with the driving semi-martingale $S_t$ in the following way: 
\[
\text{d}\lambda = \Lambda \circ dS_t.
\]
Thus, the above action can be re-written in the following form
\begin{equation*}
	\int_{t_0}^{t_1} \left(\tilde\ell(t,x,\tilde v(t,x))\circ dS_t - \langle \text{d}\lambda, C(t,x,\tilde v(t,x)) \rangle_\mathfrak{B}\right).
\end{equation*}

We may then write the constrained version of the action \eqref{FamiliarContinuumStochastic} in any of the equivalent forms
\begin{equation*}
	\int_{t_0}^{t_1}\bigg(\int_{\mathscr{D}}\ell(t,x,\tilde v(t,x))\,\mu(dx)\circ dS_t - \langle \Lambda, C(t,x,\tilde v(t,x))\circ dS_t\rangle_\mathfrak{B}\bigg),
\end{equation*}
\begin{equation*}
	\int_{t_0}^{t_1}\bigg(\int_{\mathscr{D}}\ell(t,x,\tilde v(t,x))\,\mu(dx)\circ dS_t - \langle \Lambda\circ dS_t, C(t,x,\tilde v(t,x))\rangle_\mathfrak{B}\bigg),
\end{equation*}
\begin{equation*}
	\int_{t_0}^{t_1}\bigg(\int_{\mathscr{D}}\ell(t,x,\tilde v(t,x))\,\mu(dx)\circ dS_t - \langle \text{d}\lambda, C(t,x,\tilde v(t,x))\rangle_\mathfrak{B}\bigg),
\end{equation*}
\begin{equation*}
	\int_{t_0}^{t_1}\int_{\mathscr{D}}\bigg(\ell(t,x,\tilde v(t,x))\,\mu(dx)\circ dS_t -  C(t,x,\tilde v(t,x))\,\text{d}\lambda\,\mu(dx)\bigg).
\end{equation*}
Note that in the above we have made use of the stochastic Fubini theorem. The validity of the stochastic Fubini theorem is ensured by imposing again the relevant integrability conditions.

Some constraints lend themselves very naturally to be written in an integral form. In particular, if we wish to enforce that some of the physical variables satisfy some S(P)DE, this can quite obviously be written in the form where the time integration is attached to the constraint rather than the Lagrange multiplier. However, other constraints do not necessarily lend themselves to this form. In this case, it may be more natural to write the constraint in a form where the time integration is considered as attached to the Lagrange multiplier and it may appear more intuitive to write the action in a form which includes `$\text{d}\lambda$'. See section \ref{PressureSection} for an example. 

\begin{remark}\label{MultiplierCompatibilityRemark}
We must be careful with semantics in the stochastic case. The object $\Lambda$ is the Lagrange multiplier, even though it may be tempting to refer to $\lambda$ or even $\text{d}\lambda$ as such. Furthermore, when we say that a Lagrange multiplier is compatible with a semi-martingale $S_t$, we mean that $\lambda$ is compatible with $S_t$. This is despite the fact that, formally, $\Lambda$ represents the Lagrange multipliers and may not be compatible with $S_t$.
\end{remark}

\section{A Particular Case}\label{SpecificCaseSection}

\subsection{Stochastic Advection by Lie Transport (SALT)}\label{SALTSection}

Let us assume, in some spatial domain $\mathscr{D}$, we label fluid elements by $x_0$ in an initial state, and at time $t$ these points have moved into a new configuration $x_t$. We can then define a curve, $g_t$, on the manifold of diffeomorphisms parametrized by time which is such that $x_t = g_tx_0$. In the case of deterministic fluid dynamics, we have that the Eulerian velocity field $u$ is given by $\dot x_t = \dot g_tx_0 = (u_t\circ g_t)x_0$, where the dot represents a derivative in time. We can think of $u$ as being the vector field which is driving the flow $g_t$. Following \cite{Holm2015}, suppose we want to stochastically perturb the vector field $u$.\footnote{Perturbing $u$ in this way ensures that the resulting equations feature transport noise, and is equivalent to making the assumption (from the outset) that our fluid particles follow stochastic trajectories.} We can do this by assuming that $x_t$ is a solution of
\begin{align*}
	\text{d}x_t = \text{d}g_tx_0 &= (u_t\circ g_t)x_0\,dt + \sum_{i=1}^\infty (\xi_i\circ g_t)x_0\circ dW_t^i \\
	&= u_t(x_t)\,dt + \sum_{i=1}^\infty \xi_i(x_t)\circ dW_t^i.
\end{align*}
Observe that this expression shows that the process $x_t$ is compatible with a driving semi-martingale $S_t$ given by
\begin{equation}\label{SALTDrivingSemimartingale}
    S_t = (S_t^0,S_t^1,\dots) = (t,W^1_t,W^2_t,\dots) = (t, W_t^i\ i\geq 1).
\end{equation}
 Furthermore, since this holds for any $x_0\in\mathscr{D}$, we have
\begin{equation*}
    \text{d}g_t(\cdot) = (u_t\circ g_t)(\cdot)\,dt + \sum_i (\xi_i\circ g_t)(\cdot)\circ dW_t^i,
\end{equation*}
and hence $g_t$ is also compatible with $S_t$.

Suppose $a_t:\mathscr{D}\to\mathscr{D}$ is invariant under the flow, meaning that it is an advected quantity. Then
\begin{equation*}
	a_0(x_0) = a_t(x_t) = (a_t\circ g_t)x_0 = (g^*_ta_t)x_0,
\end{equation*}
and hence, by an application of the stochastic Kunita-It\^o-Wentzell formula \cite{DeLeon2019},
\begin{equation}\label{SALTAdvection}
\begin{aligned}
	0 &= \text{d}a_0(x_0) = \text{d}(a_t\circ g_t)x_0 \\
	&= \text{d}(g^*_ta_t)x_0 \\
	&= g_t^*(\text{d}a_t(x_0) + \mathcal{L}_ua_t(x_0)\,dt + \sum_i \mathcal{L}_{\xi_i}a_t(x_0)\circ dW_t^i )\\
	&= \text{d}a_t(x_t) + \mathcal{L}_ua_t(x_t)\,dt + \sum_i \mathcal{L}_{\xi_i}a_t(x_t)\circ dW_t^i \\
	&\coloneqq \text{d}a_t(x_t) + \mathcal{L}_{\text{d}x_t}a_t.
\end{aligned}
\end{equation}
where by $g^*_ta_t$ we mean the pullback of $a_t$ by $g_t$ and we have denoted the Lie derivative by $\mathcal{L}$. We have introduced a new notation in the final line of \eqref{SALTAdvection} for a more convenient way of writing the Lie derivative terms. Namely, we absorb the notation for temporal integration into the vector field with which we are taking a Lie derivative with respect to:
\begin{equation*}
    \mathcal{L}_{\text{d}x_t}a_t \coloneqq \mathcal{L}_ua_t(x_t)\,dt + \sum_i \mathcal{L}_{\xi_i}a_t(x_t)\circ dW_t^i.
\end{equation*}
We can now give a precise definition of the constraint used in \cite{Holm2015} to introduce stochastic noise. Taking $\tilde\ell$ to be the Lagrangian by which a deterministic motion equation is derived, we can then derive our stochastic equation by applying Hamilton's principle to the action integral 
\begin{equation}\label{SALTAction_NoS}
	\int_{t_0}^{t_1}\tilde\ell(u,a_t)\,dt - \langle \Lambda,\text{d}(g^*_ta_t)\rangle
\end{equation}
which will enable us to recover the motion equations where advected quantities follow some stochastic path. Note that by enforcing $\text{d}(g^*_ta_t)=0$, we have made no assumptions on precisely how to perturb the vector field $u$. This is because in \eqref{SALTAction_NoS}, the notation we have used does not specify what the driving semi-martingale is. Should we want to obtain equations as derived in \cite{Holm2015}, we would need to consider \eqref{SALTAction_NoS} together with \eqref{SALTAdvection}, or we define our action integral instead by
\begin{equation}\label{SALTAction_S}
	A(u_t,a_t) \coloneqq \int_{t_0}^{t_1}\tilde\ell(u,a_t)\,dt - \langle\Lambda,\text{d}a_t + \mathcal{L}_ua_t\,dt + \sum_i\mathcal{L}_{\xi_i}a_t\circ dW_t^i\rangle_\mathfrak{X},
\end{equation}
where we have included the terms coming from the driving semi-martingale. Note that in practice, $L$ takes the form of a spatial integral. We define
\begin{equation*}
	\tilde\ell(u,a_t) \coloneqq \int_\mathscr{D}\ell(u,a_t)\,\mu(dx),
\end{equation*}
where $\mu$ and $\mathscr{D}$ are as defined earlier. 
Recall that we denote the stochastic derivative of $a_t$ with respect to the driving semi-martingale $S_t$ by $A_t$, this means that we have a relation $$ a_t = a_0 + \int_0^t A^0_s\,ds + \sum_{i=1}^\infty \int_0^t A^i_s\circ dW_s^i.$$ The action integral can thus be re-written as
\begin{equation*}
	A(u_t,a_t) = \int_{t_0}^{t_1}\int_\mathscr{D} \ell(u,a_t)\,\mu(dx)\,dt - \Lambda (A^0_t+\mathcal{L}_ua_t)\,\mu(dx)\,dt - \Lambda \sum_i(A^i_t +\mathcal{L}_{\xi_i}a_t)\,\mu(dx)\circ dW_t^i,
\end{equation*}
and hence has the form \eqref{FamiliarContinuumStochastic}, with $S_t$ as defined above, $\mu$ as the Lebesgue measure, and

\begin{align*}
	\ell(u,\xi_i,a_t,\Lambda) &= (\ell(u,a_t) - \Lambda A_t^0 - \Lambda \mathcal{L}_ua_t, -\Lambda A_t^1 - \Lambda\mathcal{L}_{\xi_1}a_t, -\Lambda A_t^2 - \Lambda\mathcal{L}_{\xi_2}a_t,\dots )\\
	&= (\ell(u,a_t)-\Lambda A_t^0 - \Lambda \mathcal{L}_ua_t , -\Lambda A_t^i-\Lambda \mathcal{L}_{\xi_i}a_t).
\end{align*}

By observing the action which we have defined above, one notices that we consider the time integration to be affiliated with the constraint rather than with the Lagrange multiplier. If instead we consider the Lagrange multiplier, $\Lambda$, to be the stochastic derivative of some process $\lambda$ with respect to $S_t$, then we may write this action integral in an equivalent form. Indeed, since $g_t$ and $a_t$ are compatible with $S_t$, we may define $G^*_tA_t$ to be the stochastic derivative of $g^*_ta_t$ and hence the action \eqref{SALTAction_NoS} can be written as
\begin{equation*}
    \int_{t_0}^{t_1}\tilde\ell(u,a_t)\,dt - \langle\text{d}\lambda,G^*_tA_t\rangle.
\end{equation*}

In this case, it more intuitive to write the constraint in the form \eqref{SALTAction_NoS} or \eqref{SALTAction_S}, since our constraint is a SPDE, rather than in the alternative form above. As we will see in the following subsection, in other cases it is more intuitive to write the constraint in the alternative form.

\subsection{Lagrange multipliers and the semi-martingale pressure}\label{PressureSection}

A particular advected quantity of interest is the volume element, and denote this by $D d^3x$. Note that $D$, which may be described as the fluid density, is the determinant of the Lagrange-to-Euler map. We will slightly abuse notation in the advection equation by also using $D$ to denote the whole volume element which acts as a measure of volume in our problem. Enforcing $D=1$ restricts us to incompressible flow. Regarding $D$ as a volume form\footnote{The Lie derivative of $D$ with respect to some vector field $u$ is given by ${\rm div}(Du)$.}(dropping the $d^3x$ notation), $D$ advects according to
\begin{equation}
    \text{d}D+\nabla\cdot (Du\,dt + \sum_iD\xi_i\circ dW_t^i) =0,
\end{equation}
and from this we observe that $D=1$ is equivalent to incompressibility. This is true as the relationship \[\int_0^t \nabla\cdot u_s\,ds + \sum_i\int_0^t \nabla\cdot\xi_i\circ dW_t^i =0,\] is equivalent to $\nabla\cdot\xi_i=0$ for each $i$ and $\nabla\cdot u =0$ by the uniqueness of the Doob-Meyer decomposition theorem \cite{Karatzas1998}.

The pressure, $p$, can be considered as the Lagrange multiplier which enforces the constraint $D=1$. Crucially, in our stochastic framework presented in section \ref{IntroductingStochasticity}, we can immediately see that $p$ must be compatible with the driving semi-martingale. This allows us to identify the correct form in which the pressure term appears in our motion equation, and moreover provides a rigorous framework for understanding why it must have this form. Suppose we have a physical system which corresponds to some action \eqref{StochasticTypicalAction}, and we wish to place the additional constraint of incompressibility. We are then computing the critical points of the action integral defined by

\begin{equation}
    \int_{t_0}^{t_1}\left(\tilde\ell(t,v(t))\circ dS_t-\langle \text{d}p,D-1\rangle \right),
\end{equation}

which we may re-write using the compatibility of $p$:
\begin{equation*}
    \int_{t_0}^{t_1}\tilde\ell(u,a_t)\circ dS_t -\langle P,D-1\rangle\circ dS_t.
\end{equation*}

If we want to impose this constraint on systems featuring SALT, we want to minimize action integrals of the form
\begin{equation*}
	\int_{t_0}^{t_1}\tilde\ell(u,a_t)\,dt -\langle \text{d}p,D-1\rangle - \langle \Lambda,\text{d}a_t+ \mathcal{L}_ua_t\,dt + \sum_i\mathcal{L}_{\xi_i}a_t\circ dW_t^i\rangle.
\end{equation*}

\begin{remark}
    The above action features two constraints enforced by Lagrange multipliers. Notice that we have used notation where the time integration is considered with the Lagrange multiplier, $P$, in one case and with the constraint in the other case. As mentioned above, we could write these in an equivalent form. In this case, it is more intuitive to display them as we have since, upon observing the right hand side of the inner product $\langle\cdot,\cdot\rangle$, it is immediately obvious what the physical meaning of the constraint is.
\end{remark}

Following \cite{Holm2015}, we can now derive an Euler-Poincar\'e equation for incompressible fluid equations with stochastic Lie transport. For the following theorem we maintain the notation which we have been using throughout this paper, in particular $a_t$ denotes the set of all advected quantities of which the volume element, $D$, is one.

    \begin{theorem}[Stochastic continuum Euler-Poincar\'e theorem with constraints]\label{EulerPoincareTheorem}
		Define an action integral by
		\begin{align*}
			S &= \int_{t_0}^{t_1}\tilde\ell(u,a_t)\,dt -\langle \text{d}p,D-1\rangle - \langle \Lambda,\text{d}a_t+ \mathcal{L}_{\text{d}x_t}a_t\rangle \\
			&= \int_{t_0}^{t_1} \tilde\ell(u,a_t)\,dt - \langle \text{d}p,D-1\rangle - \langle\Lambda,\text{d}a_t+\mathcal{L}_ua_t\,dt\rangle + \sum_i\langle \Lambda\diamond a_t, \xi_i(x)\rangle\circ dW^i_t,
		\end{align*}
		where $u\in\mathfrak{X}$ is the fluid velocity and $a_t$ represents the collection of advected quantities, including the volume element $D$, in $V$. Furthermore, $P$ (where $\text{d}p = P^0\,dt + \sum_i P^i\circ dW_t^i$), and $\Lambda$ are Lagrange multipliers in $V^*$. 
		An application of Hamilton's principle, $\delta S = 0$, leads to the following Euler-Poincar\'e equations
		\begin{equation}\label{EulerPoincare}
			\text{d}\frac{\delta \tilde\ell}{\delta u} +\mathcal{L}_{\text{d}x_t}\frac{\delta \tilde\ell}{\delta u} = \frac{\delta \tilde\ell}{\delta a_t}\diamond a_t\,dt- \text{d}p\diamond D,\quad \text{d}a_t+\mathcal{L}_{\text{d}x_t}a_t = 0,\quad \text{and}\quad D=1.
		\end{equation}
		In other words, $(u,a, D, P)$ is a critical path for the action $S$ if and only if the quadrouple satisfies the Euler-Poincar\'e equations \eqref{EulerPoincare}.
	\end{theorem}

    \begin{remark}
       Using the notation introduced in section \ref{StochasticContinuumDynamicsSection}, $(u,a,D,P)\in\mathfrak{B}=\bigotimes_{i=1}^4 \mathfrak{B}^i$, where $\mathfrak{B}^1 = \mathfrak{X}$, $\mathfrak{B}^2 = V$, $\mathfrak{B}^3 = V$, and $\mathfrak{B}^4 = V^*$.
    \end{remark}
	\begin{remark}
	   The second form of the action integral above features pairings $\langle\cdot,\cdot\rangle$ of two types: between elements of vector spaces and their duals, and between vector fields and their duals. These pairings are connected through the operator `$\diamond$' as in \cite{Holm2015}\footnote{For $\Lambda\in V^*$ and $a\in V$, the operator `$\diamond$' is defined such that $\Lambda\diamond a$ is the unique element in $\mathfrak{X}^*$ such that $\langle \Lambda\diamond a,\xi\rangle_\mathfrak{X} \coloneqq \langle \Lambda, - \mathcal{L}_\xi a\rangle_V$ for $\xi\in\mathfrak{X}$, see \cite{Holm2015} for further details.}.
	\end{remark}
	
	\begin{proof}
		We take variations of our action integral as follows
		\begin{align}
			\delta u&:\quad \frac{\delta \tilde\ell}{\delta u}+\Lambda\diamond a_t = 0, \label{Variation_u}\\
			\delta\Lambda &:\quad \text{d}a_t+\mathcal{L}_{\text{d}x_t}a_t=0,\label{Variation_lambda}\\
			\delta a_t&:\quad \frac{\delta \tilde\ell}{\delta a_t}\,dt + \text{d}\Lambda - \mathcal{L}_{\text{d}x_t}^T\Lambda = 0 \label{Variation_Q},\quad \text{for }a_t\neq D\\
			\delta D&:\quad \frac{\delta \tilde\ell}{\delta D}\,dt - \text{d}p + \text{d}\Lambda - \mathcal{L}_{\text{d}x_t}^T\Lambda = 0, \label{Variation_tildeq}\\
			\delta p&:\quad D-1=0. \label{Variation_p}
		\end{align}
		Notice now that equations \eqref{Variation_lambda} and \eqref{Variation_p} immediately give us two of the Euler-Poincar\'e equations. The remaining variational equations combine to give the motion equation for our system as follows. For an arbitrary vector field, $\eta$, we have
		\begin{equation*}
			\bigg\langle\text{d}\frac{\delta \tilde\ell}{\delta u} - \frac{\delta \tilde\ell}{\delta a_t}\diamond a_t\,dt,\eta\bigg\rangle = \bigg\langle -\text{d}\Lambda\diamond a_t - \Lambda \diamond\text{d}a_t - \frac{\delta \tilde\ell}{\delta a_t}\diamond a_t\,dt,\eta\bigg\rangle,
		\end{equation*}
		where we have used equation \eqref{Variation_u}. Making use of equations \eqref{Variation_lambda}, \eqref{Variation_Q}, and \eqref{Variation_tildeq} then gives
		\begin{align*}
			\bigg\langle\text{d}\frac{\delta \tilde\ell}{\delta u} - \frac{\delta \tilde\ell}{\delta a_t}\diamond a_t\,dt,\eta\bigg\rangle &= \langle -\mathcal{L}_{\text{d}x_t}^T\Lambda \diamond a_t - \text{d}p\diamond D + \Lambda \diamond \mathcal{L}_{\text{d}x_t}a_t ,\eta\rangle \\
			&= \langle \Lambda , \mathcal{L}_{\text{d}x_t}\mathcal{L}_\eta a_t - \mathcal{L}_\eta\mathcal{L}_{\text{d}x_t}a_t\rangle - \langle \text{d}p\diamond D ,\eta\rangle \\
			&=  \langle \Lambda\diamond a_t, \text{ad}_{\text{d}x_t}\eta\rangle - \langle \text{d}p\diamond D ,\eta\rangle\\
			&= \langle \text{ad}^*_{\text{d}x_t}(\Lambda\diamond a_t),\eta\rangle - \langle \text{d}p\diamond D ,\eta\rangle \\
			&= \bigg\langle -\mathcal{L}_{\text{d}x_t}\frac{\delta \tilde\ell}{\delta u} - \text{d}p\diamond D,\eta\bigg\rangle.
		\end{align*}
		Since our choice of the vector field $\eta$ was arbitrary, this completes our proof.
	\end{proof}
	
The action integral, $A$, used in the above theorem for incompressible SALT equations is of the form \eqref{FamiliarContinuumStochastic}, with $S_t$ as defined in section \ref{SALTSection} and

\begin{equation*}
	\ell = (\ell(u,a_t)-P_0(D-1)-\Lambda A_t^0 - \Lambda \mathcal{L}_ua_t , -P_i(D-1)-\Lambda A_t^i-\Lambda \mathcal{L}_{\xi_i}a_t),
\end{equation*}
where $\ell$ has infinitely many components and we have represented these by indexing them by $i$.

\begin{remark}
   Should the action integral be written in such a way that the integrand contains terms of the form $\text{d}\lambda$, where $\lambda$ is compatible with the driving semi-martingale, then it is technically improper to refer to a variational derivative with respect to $\lambda$. Instead the action should be considered in its equivalent form \eqref{FamiliarContinuumS}, in which case we perform variational derivatives with respect to $\Lambda$ (the stochastic derivative of $\lambda$ with respect to $S_t$). Nonetheless, when considering the pressure term in section \ref{PressureSection}, to ease notation we will refer to variations with respect to $p$ rather than $P$, and we here define these variations to be equivalent. 
\end{remark}
	
\subsection{Example: A stochastic Euler equation for incompressible fluids}

As discussed in previous sections, we will derive our equations by applying Hamilton's principle to the relevant Lagrangian. Consider the Lagrangian needed to derive the deterministic Euler equations from a variational principle \cite{Arnold1966}, which can be defined by
	\begin{equation}\label{EulerLagrangian}
		\tilde\ell(u,D)=\int_\mathscr{D}\frac{1}{2}D|u|^2\,dx,\quad D=\frac{\rho}{\rho_0},
	\end{equation}
	and now consider the constrained action integral given by
	\begin{equation}\label{EulerAction}
		A(u,D,p,\Lambda) = \int_{t_0}^{t_1}\bigg(\tilde\ell(u,D)\,dt - \langle \text{d}p, D-1\rangle -\langle\Lambda,\text{d}D+\mathcal{L}_{\text{d}x_t}D\rangle\bigg).
	\end{equation}
	Recall that, in our notation, $a$ denotes the set of all advected quantities, including $D$. However, in this case, $D$ is the only advected quantity we consider. Again, by unravelling $\text{d}p$, $\text{d}D$, and $\mathcal{L}_{\text{d}x_t}D$, we can verify that this integral is of the form \eqref{StochasticTypicalAction}.

	\begin{remark}
	   In the deterministic case, the constraint and Lagrange multiplier for incompressibility are incorporated into the Lagrangian, which becomes \[\tilde\ell(u,D,p) = \int_\mathscr{D}\left(\frac{1}{2}D|u|^2 - p(D-1)\right)\,dx.\]
	   Since this Lagrangian can be used to derive the deterministic incompressible Euler equation; when attempting to derive the incompressible SALT Euler equation, it is tempting to use the same Lagrangian in the action integral \eqref{SALTAction_S}. Note that this is insufficient, since SALT introduces a stochastic noise which defines the driving semi-martingale of the action integral (the addition of SALT can be thought of as changing the driving semi-martingale from $t$ to $(t,W_t^1,W_t^2,\dots)$). Thus we need to take care to ensure that the Lagrange multiplier is compatible with the driving semi-martingale (rather than with $t$), and hence it is essential to write it outside of the deterministic Lagrangian. The concept of a driving semi-martingale allows us to correctly characterise the pressure at the level of the Lagrangian, rather than observing in the equation that it must have an additional stochastic part. Without the framework defined in this paper, from the variational perspective it is not clear why the pressure cannot manifest itself as `$p\,dt$' in the equation of motion.
	\end{remark}
	
	We thus apply our Euler-Poincar\'e equations \eqref{EulerPoincare} to the action integral \eqref{EulerAction}, which gives us
	\begin{equation*}
		\text{d}u + (\text{d}x_t\cdot\nabla)u + (\nabla \text{d}x_t)\cdot u = \frac{1}{2}\nabla |u|^2\,dt - \nabla\text{d}p,
	\end{equation*}
	where $\text{d}x_t$ is as defined earlier. Notice that the equations \eqref{EulerPoincare} contain the terms $\frac{\delta \tilde\ell}{\delta a_t}\diamond a_t\,dt$ and $\text{d}p \diamond D$. In this case, we have
	\begin{align*}
		\frac{\delta \tilde\ell}{\delta a_t}\diamond a_t\,dt &= D\nabla\bigg(\frac{\delta \tilde\ell}{\delta D}\bigg)\,dt, \\
		\text{d}p \diamond D &= D\nabla \text{d}p,
	\end{align*}
	see \cite{Holm2015} for details. Continuing with our derivation of the Euler equations, notice that the deterministic part of $(\nabla \text{d}x_t)\cdot u$ cancels with $\frac{1}{2}\nabla |u|^2\,dt$ to give
	\begin{equation*}
		\text{d}u + (\text{d}x_t\cdot\nabla)u + \sum_{i=1}^\infty (\nabla \xi_i)\cdot u\circ dW^i_t =  - \nabla\text{d}p,
	\end{equation*}
	which is obviously equivalent to the `stochastic advection by Lie transport' version of the Euler equations for incompressible fluids, given by
	\begin{align}
		\text{d} u + u\cdot\nabla u\,dt + \sum_{i=1}^\infty\bigg(\xi_i\cdot\nabla u+ \sum_{j=1}^N u_j\nabla\xi^j_i\bigg)\circ dW_t^i + \nabla\text{d}p &= 0, \label{EulerHolm} \\
		\nabla\cdot\text{d}x_t &=0, \label{DxIncompressible}
	\end{align}
	where condition \eqref{DxIncompressible} is equivalent to incompressibility as previously discussed.
	
	\begin{remark}
		Notice that $\nabla v$ is a second order tensor for any vector $v$, and thus by $(\nabla v)\cdot u$ we mean $\sum_{j}u_j\nabla v_j$.
	\end{remark}
	Now, we consider equation \eqref{EulerHolm} in its integral form and take divergence of this to give
	
	\begin{align*}
			\nabla\cdot(u_t-u_0) + \sum_{k=1}^\infty \int_0^t (\Delta \xi_k)\cdot u &+ \sum_{i,j=1}^N\bigg(\frac{\partial \xi_k^i}{\partial x_j}\frac{\partial u_i}{\partial x_j}+\frac{\partial\xi_k^i}{\partial x_j}\frac{\partial u_j}{\partial x_i}\bigg)\circ dW_s^k \\
			&+\int_0^t \sum_{i,j=1}	^N\frac{\partial u_i}{\partial x_j}\frac{\partial u_j}{\partial x_i}\,ds +\Delta p_t - \Delta p_0 = 0.
	\end{align*}
	Incompressibility then gives that $p$ satisfies the following
	
	\begin{align}
		\text{d}\pi &= -\sum_{i,j=1}	^N\frac{\partial u_i}{\partial x_j}\frac{\partial u_j}{\partial x_i}\,dt - \sum_{k=1}^\infty \bigg( (\Delta \xi_k)\cdot u + \sum_{i,j=1}^N\bigg(\frac{\partial \xi_k^i}{\partial x_j}\frac{\partial u_i}{\partial x_j}+\frac{\partial\xi_k^i}{\partial x_j}\frac{\partial u_j}{\partial x_i}\bigg)\bigg)\circ dW_t^k, \\
		p &= \Delta^{-1}\pi.
	\end{align}
	
	Note that if we considered the Lagrange multiplier of $D-1$ in the derivation to be simply $p\,dt$ or similar, even if we considered $p$ to be a stochastic process then this would be insufficient since at the level of the equation it would imply that a sum of Lebesgue integrals is equal to a stochastic integral (after taking divergence of each side of the equation). This example provides an explicit illustration of why the Lagrange multiplier needs this more general form we have presented in this paper.

\subsection{Example: A stochastic rotating shallow water equation for compressible fluids}
Take $u$ to be the fluid velocity field, $f$ to be the Coriolis parameter, $b$ to be the bottom topography of our domain $\mathscr{D}$, and $\eta$ to be its total depth. We define $R$ to be such that $\nabla\times R = f\hat{z}$, where $\hat{z}$ is the unit vector in the vertical direction.

The deterministic rotating shallow water equations (rSWE), are the following nonlinear partial differential equations \cite{MPEBook}
\begin{align*}
    \varepsilon\left(\frac{\partial u}{\partial t} + u\cdot\nabla u\right) + f\,\hat{z}\times u + \nabla k &= 0 \\
    \frac{\partial\eta}{\partial t} + \nabla\cdot(\eta u) &= 0,
\end{align*}
where $k \coloneqq (\eta - b) / \varepsilon\mathcal{F}$. The momentum equation can be shown to have the following curl form:
\begin{equation*}
    \partial_t(\varepsilon u + R) - u \times \rm{curl}(\varepsilon u + R) + \nabla\left(k + \frac{\varepsilon}{2}|u|^2\right) = 0.
\end{equation*}
The deterministic rSWE may be deduced by applying Hamilton's principle to the Lagrangian defined by
\begin{equation*}
    \tilde\ell = \int_\mathscr{D}\frac{\varepsilon}{2} \eta|u|^2 + \eta\,u\cdot R - \frac{(\eta - b)^2}{2\varepsilon\mathcal{F}}\,d^2x,
\end{equation*}
see \cite{MPEBook}. If we augment this with the stochastic advection constraint as found in Section \ref{SALTSection}, the resulting action integral is stochastic and has the form we have presented in this paper and we may deduce the stochastic rSWE featuring SALT. For the Lagrangian defined above, we may apply the stochastic Euler-Poincar\'e theorem as stated in \cite{Holm2015} which implies that
\[
\text{d}\frac{\delta \tilde\ell}{\delta u} + \mathcal{L}_{\text{d}x_t}\frac{\delta \tilde\ell}{\delta u} = \eta\nabla\frac{\delta \tilde\ell}{\delta \eta}\,dt.
\]
If we set $m \coloneqq \delta\tilde\ell / \delta u$, we see that
\begin{align*}
    \text{d}\left(\frac{m}{\eta}\right) + \mathcal{L}_{\text{d}x_t}\left(\frac{m}{\eta}\right) &= \frac{1}{\eta}\text{d}m - m\text{d}\left(\frac{1}{\eta}\right) + \frac{1}{\eta}\mathcal{L}_{\text{d}x_t}m + m\mathcal{L}_{\text{d}x_t}\left(\frac{1}{\eta}\right) \\
    &= \frac{1}{\eta}\left(\text{d}m - \frac{m}{\eta}\left(\text{d}\eta+\mathcal{L}_{\text{d}x_t}\eta\right)+\mathcal{L}_{\text{d}x_t}m\right) \\
    &= \frac{1}{\eta}(\text{d}m + \mathcal{L}_{\text{d}x_t}m),
\end{align*}
and hence we have
\begin{equation*}
    \text{d}(\varepsilon u + R) + \mathcal{L}_{\text{d}x_t}(\varepsilon u + R) = \nabla\left(\frac{\varepsilon}{2}|u|^2 + u\cdot R - k \right)dt.
\end{equation*}
This is equivalent to
\begin{equation*}
    \varepsilon\text{d}u - \text{d}x_t \times \rm{curl}(\varepsilon u + R) + \sum_i\nabla\big(\xi_i\circ dW_t^i\cdot (\varepsilon u + R)\big) = \nabla \left(\frac{\varepsilon}{2}|u|^2 - k\right)dt,
\end{equation*}
which is the stochastic rSWE in curl form.

\subsection{Example: Stochastic wave current interaction}

As introduced in \cite{holm2020variational}, the action integral for SALT can be coupled to a stochastic wave action which introduces a nonlinear wave propagation. Maintaining notation from previous section, we introduce conjugate wave variables $(p,q)$\footnote{By $p$ here we do not mean the pressure.} and define a coupled action integral in terms of the wave Hamiltonian by
\begin{equation}\label{SWCIAction}
    \begin{aligned}
        A(u,a,\Lambda,p,q) = \int_{t_0}^{t_1}&\tilde\ell(u,a)\,dt + \int_{t_0}^{t_1}\left\langle\Lambda,\text{d}a+\mathcal{L}_{\text{d}x_t}a\right\rangle_V \\
        &- \int_{t_0}^{t_1}\left\langle p\diamond q, \text{d}x_t\right\rangle_\mathfrak{X} + \int_{t_0}^{t_1}\Big(\left\langle p,\text{d}q \right\rangle_V - \text{d}\mathcal{J}(p,q)\Big),
    \end{aligned}
\end{equation}
where the first integral corresponds to the deterministic fluid action, the second to the stochastic advection constraint, the third to a form of minimal coupling, and the fourth to the phase-space wave Lagrangian. The stochastic wave Hamiltonian is defined by
\begin{equation*}
    \text{d}\mathcal{J}(p,q) \coloneqq \mathcal{H}(p,q)\,dt + \mathcal{K}(p,q)\circ dB_t,
\end{equation*}
and $\text{d}x_t$ is defined as in Section \ref{SALTSection}. Here, we choose $B_t$ to be a Brownian motion independent to each $W_t^i$ to ensure that there is no unwanted interaction between the stochastic noise in the fluid transport and that in the wave Hamiltonian. The drift part of $\text{d}\mathcal{J}(p,q)$ is some deterministic wave Hamiltonian which is suitable to the problem, whilst the diffusion part is \emph{chosen} to be a pairing between a vector field $\sigma(x)$ and the wave momentum map:
\begin{equation*}
    \mathcal{K}(p,q)\circ dB_t = \langle p\diamond q, \sigma(x) \rangle_\mathfrak{X} \circ dB_t.
\end{equation*}
This choice means that the stochastic part of the wave Hamiltonian enforces an additional stochastic transport of the wave properties by the vector field martingale $\sigma(x)\circ dB_t$.

The action integral \eqref{SWCIAction} can clearly be written in the form \eqref{FamiliarContinuumS}, where the driving semi-martingale is
\[
S_t = (t,B_t,W_t^1,W_t^2,W_t^3,\dots),
\]
and our variables are compatible with $S_t$ as required.
\begin{remark}[Duelling semi-martingales]
    Our action integral for stochastic wave current interaction contains a stochastic noise in the fluid transport as well as an independent noise in the wave Hamiltonian. We can `switch off' either (or both) of these stochastic noises by setting $\mathcal{K}(p,q)=0$ or (and) $\xi_t=0$ for each $i$. Since $B_t$ is independent of each $W_t^i$, switching off one of these noises will not interfere with the function of the other. Nonetheless, despite this independence, the resulting system of equations is a coupled system and therefore the wave variables will feel the effects of $W_t^i$ and the fluid variables the effect of $B_t$. Therefore, whilst it may be tempting to believe that the wave variables need only be compatible with $(t,B_t)$ and the fluid variables with $(t,W_t^i\ i\geq 1)$, in fact both the wave and fluid variables will need to be compatible with the \emph{full} driving semi-martingale $S_t$ as defined above. Furthermore, any additional constraint placed on this action via a Lagrange multiplier will require that the Lagrange multiplier is compatible with the full driving semi-martingale, even if that constraint is only physically meaningful to the fluid (or wave) dynamics in isolation.
\end{remark}

\section{Summary and future work}

In this work we have introduced a rigorous framework for applying the principle of least action to stochastic action integrals. We have shown that a stochastic version of the fundamental Lemma of the calculus of variations holds, and that it is therefore permissible to take variations of stochastic integrals. We have also shown that the resulting relations obtained by varying the action integral have a similar structure to the deterministic picture. Namely, we obtain a stochastic equation of motion depending on variational derivatives of the Lagrangian, where these variational derivatives are identical in definition to the classical framework.

Crucially, the new stochastic framework clarifies the required form of Lagrange multipliers within a semi-martingale driven variational principle. This in turn clarifies how one can formulate stochastic variational principles to ensure that the resulting equations are well-defined mathematically and have the desired physical interpretation. By understanding the role of the driving semi-martingale, one can keep track of the distinct forms of time integration within the system and ensure that all are being considered correctly within each term of the action integral.

Many further problems have arisen from the framework we have introduced in this paper. Perhaps most glaringly, the details of an Euler-Poincar\'e reduction for a semi-martingale driven action integral are non-trivial and need to be carefully studied. Furthermore, the introduction of this framework opens up the possibility of determining how a Legendre transform may connect stochastic Lagrangian mechanics of this type to stochastic Hamiltonian mechanics (this been formulated rigorously \cite{Lazaro-Cami2008}).

\section*{Funding}
During this work, the first author has been supported by an EPSRC studentship as part of the Centre for Doctoral Training in the Mathematics of Planet Earth (grant number EP/L016613/1).  The second author has been partially supported by EU project STUOD - DLV-856408.

\section*{Acknowledgements}
This work was done following the first introduction of stochastic variational principles for fluid dynamics in the seminal paper \cite{Holm2015}. It is the result of many discussions that we have had with our colleagues Darryl Holm, James Michael Leahy, Erwin Luesink, So Takao, Etienne Memin, Colin Cotter, Oana Lang. We thank them for generously exchanging ideas with us about their own related work.

\bibliographystyle{abbrv}
\bibliography{PhDProject.bib}

\end{document}